# The Influence of Surface Topography and Surface Chemistry on the Anti-Adhesive Performance of Nanoporous Monoliths

Anna Eichler-Volf,[†] Longjian Xue,[$,*] Gregor Dornberg,[&] He Chen,[‡] Alexander Kovalev,[§] Dirk Enke,[&] Yong Wang,[‡*] Elena V. Gorb,[§] Stanislav N. Gorb,[§*] Martin Steinhart[†*]

[†] Institut für Chemie neuer Materialien, Universität Osnabrück, Barbarastr. 7, 49069 Osnabrück, Germany

[$] School of Power and Mechanical Engineering; Wuhan University; South Donghu Road 8, Wuhan, Wuchang, 430072; Hubei, China

[§] Functional Morphology and Biomechanics, Zoological Institute, Kiel University, Am Botanischen Garten 9, 24118 Kiel, Germany

[&] Institut für Technische Chemie, Universität Leipzig, Linnéstr. 3-4, 04103 Leipzig, Germany

[‡] State Key Laboratory of Materials-Oriented Chemical Engineering, College of Chemical Engineering, Nanjing Tech University, Nanjing 210009, Jiangsu, P. R. China

ABSTRACT

We designed spongy monoliths allowing liquid delivery to their surfaces through continuous nanopore systems (mean pore diameter ~40 nm). These nanoporous monoliths were flat or patterned with microspherical structures a few 10 μm in diameter, and their surfaces consisted of aprotic polymer or of $TiO_2$ coatings. Liquid may reduce adhesion forces $F_{Ad}$; possible reasons include screening of solid-solid interactions and poroelastic effects. Softening-induced deformation of flat polymeric monoliths upon contact formation in the presence of liquids enhanced the work of separation $W_{Se}$. On flat $TiO_2$-coated monoliths, $W_{Se}$ was under wet conditions smaller than under dry conditions, possibly because of liquid-induced screening of solid-solid interactions. Under dry conditions, $W_{Se}$ is larger on flat $TiO_2$–coated monoliths than on flat monoliths with polymeric surface. However, under wet conditions liquid-induced softening results in larger $W_{Se}$ on flat monoliths with polymeric surface than on flat monoliths with oxidic surface. Monolithic microsphere arrays show anti-adhesive properties; $F_{Ad}$ and $W_{Se}$ are reduced by at least one order of magnitude as compared to flat nanoporous counterparts. On nanoporous monolithic microsphere arrays, capillarity ($W_{Se}$ is larger under wet than under dry conditions) and solid-solid interactions ($W_{Se}$ is larger on oxide than on polymer) dominate contact mechanics. Thus, the microsphere topography reduces the impact of softening-induced surface deformation and screening of solid-solid interactions associated with liquid supply. Overall, simple modifications of surface topography and chemistry combined with delivery of liquid to the contact interface allow adjusting $W_{Se}$ and $F_{Ad}$ over at least one order of magnitude. Adhesion management with spongy monoliths exploiting deployment (or drainage) of interfacial liquids as well as induction or prevention of liquid-induced softening of the monoliths may pave the way for the design of artificial surfaces with tailored contact mechanics. Moreover, the results reported here may contribute to better understanding of the contact mechanics of biological surfaces.

INTRODUCTION

Extensive research activities have been devoted to the design of bioinspired adhesive pads forming strong reversible contact to counterpart surfaces via arrays of fibrillar contact elements.[1,2] The design of anti-adhesive surfaces has attracted much lesser attention. Among the origins of anti-adhesive behaviour, surface topographies leading to a decrease in the actual contact area with a counterpart surface play a prominent role. The pull-off force (also referred to as adhesion force) $F_{Ad}$ between two contacting counterpart surfaces can be reduced by surface topographies that reduce the real contact area.[3] Focussing on topographic effects, anti-adhesive behaviour can be considered as topography-induced reduction of $F_{Ad}$ in comparison with flat reference surfaces. Focussing on surface chemistry, anti-adhesive behaviour can be considered as reduction of $F_{Ad}$ caused by a variation in the chemical nature of a surface while other parameters such as topography are kept unaltered. The design of anti-adhesive artificial surfaces has predominantly been inspired by plant surfaces with hierarchical topographic features.[4-10] Typically, the first hierarchical level is mimicked by arrays of microspheres with radii $r_s$ of the order of a few 10 μm or other artificial surfaces with similar topographies, whereas various synthetic approaches have been applied to implement further hierarchical structure levels.[11-18] The rational design of anti-adhesive artificial surfaces encounters two challenges. First, any rough surface will show anti-adhesive properties on rigid counterpart surfaces as contact is only formed at protrusions, resulting in real contact areas much smaller than the contour of the apparent macroscopic contact area. However, sticky and compliant surfaces can adapt to surface topographies characterized by feature sizes below a few microns.[19] Therefore, the contact between the contacting surfaces is conformal; the real contact area and $F_{Ad}$ may even be enhanced as compared to contacts between two flat surfaces. Secondly, the detection of anti-adhesive properties has remained challenging. This problem has recently been addressed by the use of sticky and compliant poly(dimethyl silioxane) (PDMS) half-spheres as probes for investigation of adhesion on weakly adhesive surfaces.[19] Thus, we could show that solid monolithic polystyrene (PS) microsphere arrays with $r_s$ values in the 10 μm range are anti-adhesive on sticky and compliant counterpart surfaces.[20] Nanoporous monolithic microsphere arrays (NMMAs) with $r_s$ values of a few 10 μm, which consisted of the block copolymer polystyrene-*block*-poly(2-vinylpyridine) (PS-*b*-P2VP), showed antiadhesive behaviour at low relative humidities of 2 % as well as at high relative humidities of 90 %.[21]

Besides the optimization of the contact topography, supply of liquid to contact interfaces is a second option for adhesion management. Artificial anti-icing surfaces were reported to rely on the immobilization of lubricants within porous scaffolds.[22-24] Moreover, supply of adhesive secretion to the contact interfaces of insect feet[25,26] through channels or sponge-like pore systems[27,28] was shown to affect the adhesive performance of the insects' attachment devices.[29-34] Only little efforts have been directed to the investigation of wet adhesion on artificial nanopatterned and/or micropatterned surfaces, such as frog-inspired solid micropillar arrays.[35] Recently, we have prepared fibrillar PS-*b*-P2VP adhesive pads that contained continuous, spongy nanopore systems with pore diameters of a few 10 nm. These fibrillar adhesive pads were designed for strong reversible adhesion and formed contact to counterpart surfaces via dense arrays of nanorod-like contact elements with diameters of a few 100 nm. We found that humidity-induced softening of the fibrillary adhesive pads[36] as well as supply of liquid through the nanopore systems to the contact interface[37] significantly increased adhesion. The latter effect was attributed to capillarity-supported formation of solid-solid contact between the nanorod-like contact elements of the fibrillar adhesive pads and rigid counterpart surfaces.

Continuous nanopore systems allowing supply of liquid to contact interfaces have, to the best of our knowledge, not been combined with anti-adhesive surface topographies such as monolithic arrays of microspheres with $r_s$ values in the 10 μm range. The influence of wet conditions on adhesion has hardly been studied for anti-adhesive surface topographies and, apart from nanoporous fibrillar adhesive pads optimized for strong adhesion,[37] for nanoporous surfaces of any topography. Finally, it has not been investigated how the adhesive properties of nanoporous surfaces under dry and wet conditions differ for different chemical compositions of the contact surface. Here we address these problems by studying adhesion on NMMAs with $r_s$ values of a few 10 μm and on macroscopically flat but likewise nanoporous PS-*b*-P2VP monoliths. Since liquids like mineral oil can be supplied to the contact interfaces through the spongy-continuous nanopore systems, adhesion could be studied comparatively under dry and wet conditions. Pore walls and outer surfaces of the samples initially consisted of the aprotic polymer P2VP. We deposited conformal titania ($TiO_2$) layers on the P2VP surfaces of some samples by atomic layer deposition (ALD)[38] in such a way that the nanoporous structure was conserved. Whereas P2VP cannot form hydrogen bonds with the crosslinked PDMS used as counterpart surface, the acidic hydrogen atoms of the $TiO_2$ hydroxyl groups can form hydrogen bonds with the backbone oxygen atoms of the PDMS. As discussed below, the contact mechanics of the nanoporous PS-*b*-P2VP monoliths can be interpreted as complex interplay of solid/solid interactions, the presence or absence or capillarity, the mechanical properties of the tested samples, and topographic effects. Adjusting these parameters may pave the way for liquid-mediated adhesion management involving supply or drainage of liquids to or from contact surfaces as well as liquid-induced contact surface softening. We show that $F_{Ad}$ and the work of separation $W_{Se}$ on monoliths containing sponge-like continuous nanopore systems with pore diameters of a few 10 nm can be varied by at least one order of magnitude.

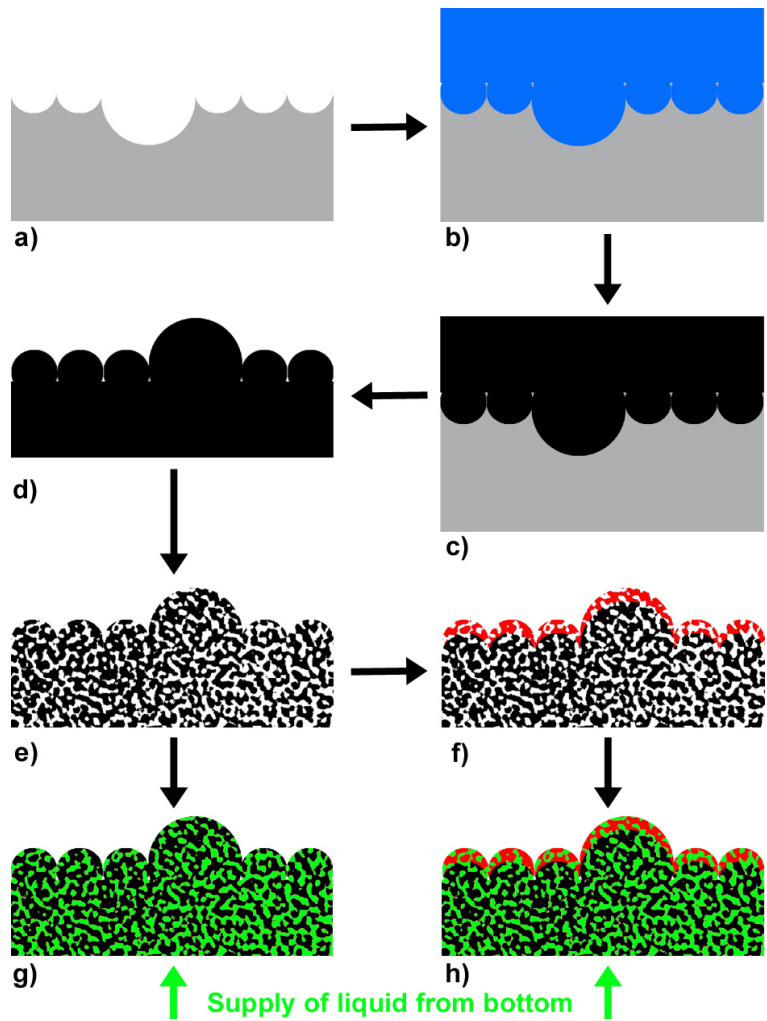

**Figure 1. Preparation of nanoporous monolithic microsphere arrays (NMMAs) and supply of liquid to the contact interface.** a) A PDMS secondary mold (grey) is prepared by replication of a PS microsphere monolayer used as primary mold. b) A solution of the block copolymer PS-*b*-P2VP (blue) is deposited into the PDMS secondary mold. c) The solvent is slowly evaporated so that a PS-*b*-P2VP monolith (black) remains in the PDMS secondary mold. d) The PDMS secondary mold is nondestructively detached. e) A spongy continuous nanopore system is generated in the PS-*b*-P2VP monolith by swelling-induced pore generation so that a NMMA is obtained. f) Optionally, the NMMA surface is modified with $TiO_2$ (red) by ALD so that a $TiO_2$-NMMA is obtained. g), h) Liquid (green) is injected into NMMAs and $TiO_2$-NMMAs from the underside opposite to the topographically structured contact interface.

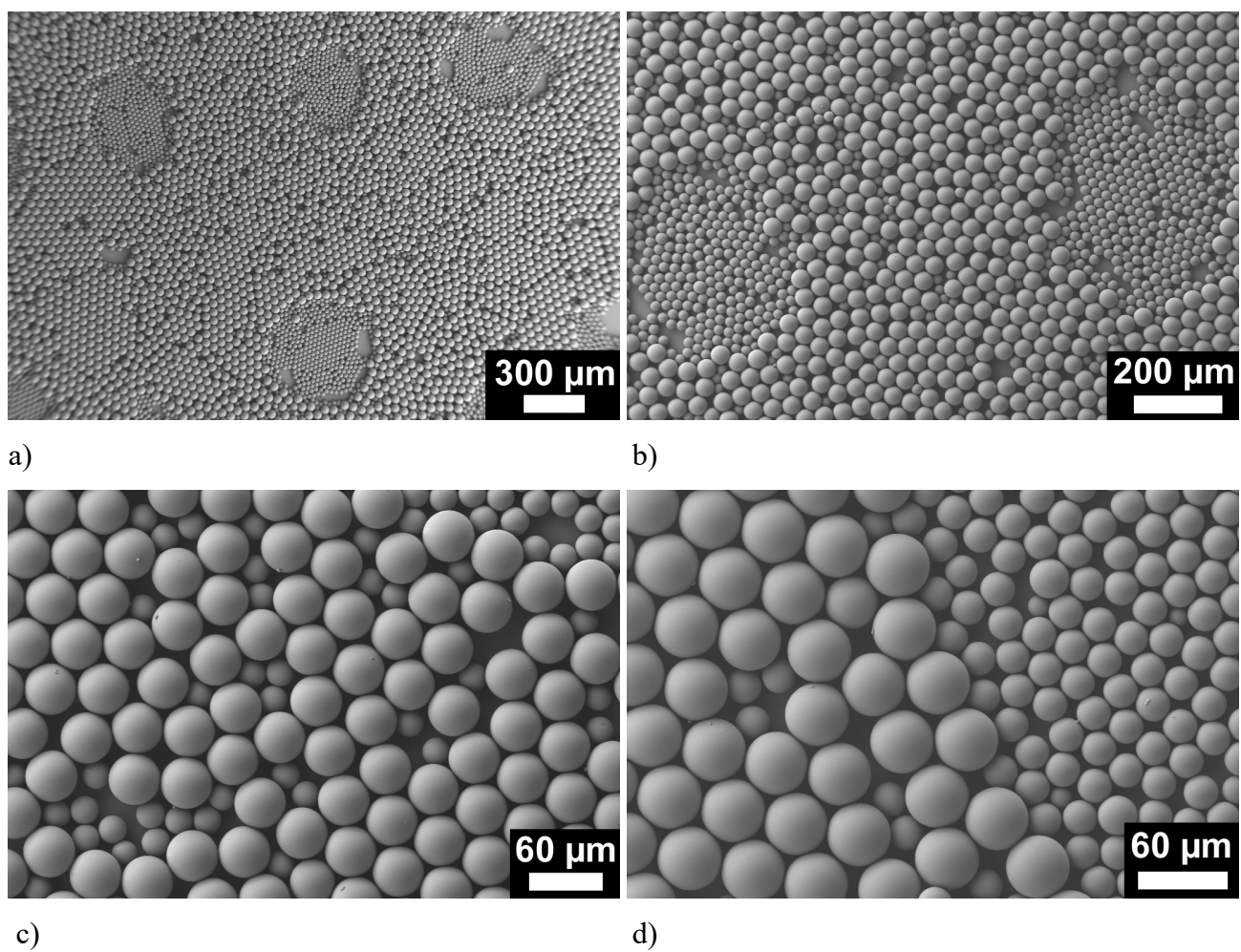

**Figure 2.** Scanning electron microscopy images of a mixed monolayer of PS microspheres with radii of 12.5 µm and 22.5 µm on a silicon wafer prepared by spin coating (cf. experimental part) used as primary mold to prepare PDMS secondary molds (cf. Figure 1a). a), b) Large-field view (image field: ~ 3 mm x ~2 mm) showing discrete domains containing PS microspheres with $r_s$ = 12.5 µm (right) surrounded by a matrix with high content of PS microspheres with $r_s$ = 22.5 µm and low content of PS microspheres with $r_s$ = 12.5 µm; c) typical matrix area with high content of PS microspheres with $r_s$ = 22.5 µm; d) boundary between a domain containing PS microspheres with $r_s$ = 12.5 µm (right) and matrix (left).

RESULTS

**Preparation of flat nanoporous PS-*b*-P2VP monoliths (FNMs), NMMAs and their TiO$_2$-coated derivatives**

Mechanically robust ~500 µm thick FNMs containing continuous spongy nanopore systems were obtained by swelling-induced pore generation [39-42] of solid films consisting of asymmetric PS-*b*-P2VP, as discussed in reference 21. As demonstrated previously, it is possible to transport liquid through these continuous pore systems.[36,40,42] NMMAs were obtained by double replication[20,43,44] of mixed monolayers of PS microspheres with radii $r_s$ of 12.5 µm and 22.5 µm (cf. reference 20 and "Materials and methods" section) that is schematically displayed in Figure 1. Since the NMMAs were faithful positive replicas of the mixed PS microsphere monolayers, which are thereafter referred to as primary molds, the latter defined the surface topography of the NMMAs. As obvious from Figure 2a, the mixed PS microsphere monolayers consist of discrete spherical to ellipsoidal domains of PS microspheres with $r_s$ = 12.5 µm extending several 100 µm surrounded by a continuous matrix domain containing a high proportion of PS microspheres with $r_s$ = 22.5 µm. The discrete spherical to ellipsoidal domains exclusively contain PS microspheres with $r_s$ = 12.5 µm (Figure 2b), whereas the continuous matrix domain contains, besides the PS microspheres with $r_s$ = 22.5 µm, a significant proportion of PS microspheres with $r_s$ = 12.5 µm (Figure 2c). The domain boundaries are, however, sharp (Figure 2d). Molding PDMS prepolymer formulation against the mixed PS microsphere monolayers used as primary molds combined with curing of the PDMS yielded PDMS secondary molds (Figure 1a). The PDMS secondary molds contained spherical cavities as negative replicas of the spin-coated discrete PS microspheres. Positive PS-*b*-P2VP replicas of the primary molds were produced in a second, nondestructive replication step involving deposition of PS-*b*-P2VP solutions onto the secondary molds (Figure 1b) and slow drying (Figure 1c). After non-destructive detachment from the PDMS secondary mold (Figure 1d), the PS-*b*-P2VP specimens were subjected to swelling-induced pore generation, as described above (Figure 1e).

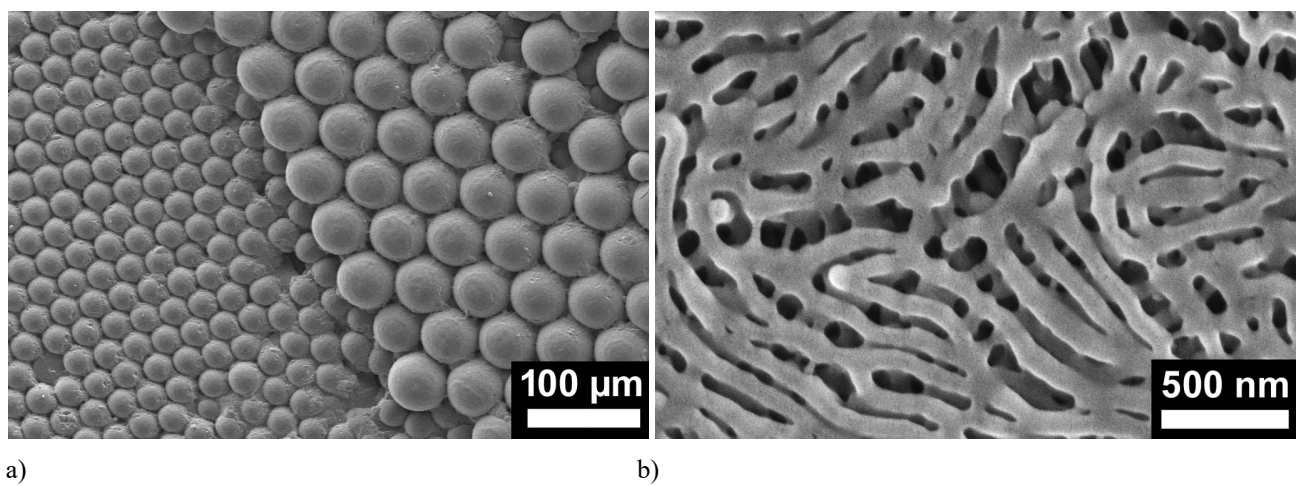

a) b)

**Figure 3.** SEM images of TiO$_2$(100)-NMMAs. a) Large-field view; b) detail.

Scanning electron microscopy (SEM) (Figure S1) as well as mercury intrusion and N$_2$-sorption (Figure S2) confirmed that swelling-induced pore generation yielded continuous nanopore systems open to the environment penetrating the entire thickness of the NMMAs. Moreover, swelling-induced pore generation did not alter the surface topography of the monolithic PS-*b*-P2VP specimens on length scales larger than the nanopore diameters. In the case of NMMAs, the surface topography obtained by double replication of the primary PS microsphere molds was conserved (Figure S1a). Closer inspection revealed that the surface of the NMMAs was nanoporous (Figure

S1b). SEM investigations of NMMA cross sections obtained by cleaving evidenced that the NMMAs were uniformly nanoporous across their entire thickness (Figure S1c,d). The mean nanopore diameter of the NMMAs amounted to about 40 nm, their specific surface area to 10 m$^2$/g and their total nanopore volume to 0.05 cm³/g (Figure S2).

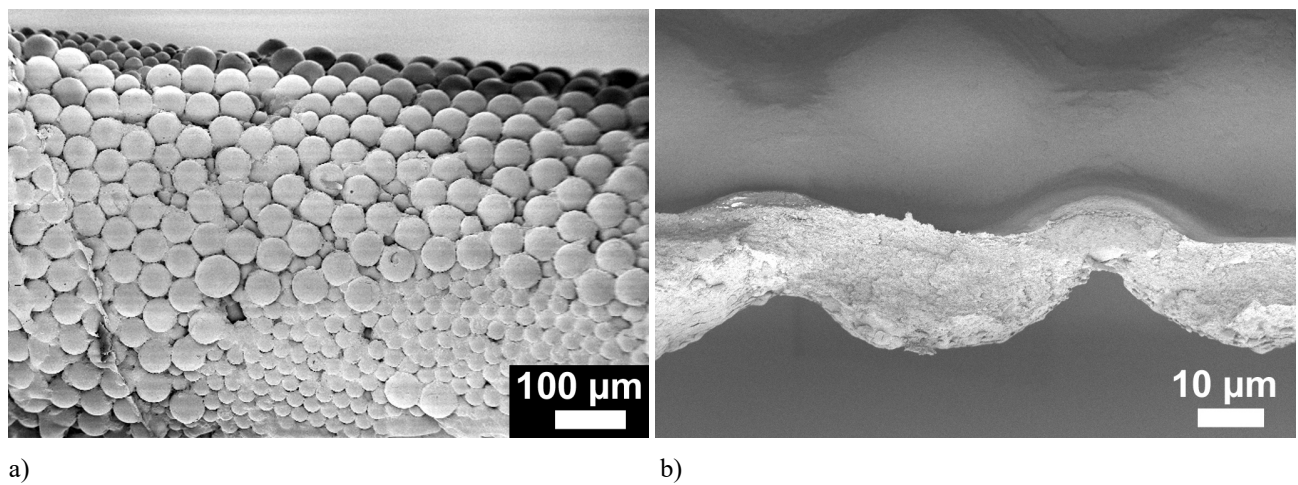

a)                            b)

**Figure 4.** SEM images of freestanding TiO$_2$ layers obtained from a TiO$_2$(60)-NMMA by extraction of the PS-*b*-P2VP. a) Top view of a freestanding TiO$_2$ layer coiled after extraction; b) cross-sectional side view of a freestanding TiO$_2$ layer.

To alter the chemical nature of their outer surface, some FNMs (thereafter referred to as TiO$_2$(60)-FNMs and TiO$_2$(100)-FNMs) as well as some NMMAs (thereafter referred to as TiO$_2$(60)-NMMAs and TiO$_2$(100)-NMMAs) were coated with TiO$_2$ by 60 or 100 low-temperature ALD cycles (Figure 1f),[38] as described below in the "Materials and Methods" section. Both the microsphere arrays defining the macroscopic topography (Figure 3a) and the nanopore systems obtained by swelling-induced pore generation (Figure 3b) were faithfully retained even after 100 ALD cycles. Treatment of TiO$_2$(60)-NMMAs with tetrahydrofuran (THF) for 0.5 h at room temperature yielded free-standing TiO$_2$ films that were free of cracks, mechanically robust and could easily be handled with tweezers (Figure 4). Even though the TiO$_2$ films coiled during drying, they faithfully mimicked the surface topography of the NMMAs.

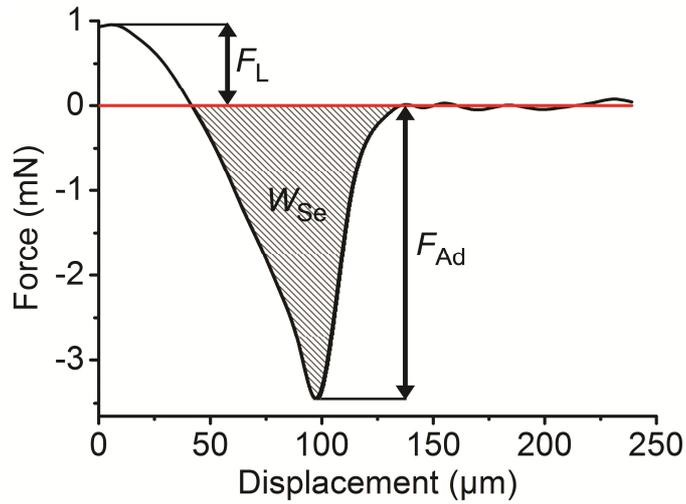

**Figure 5.** Retraction part of a force-displacement curve measured on a FNM using a PDMS half-sphere (effective elastic modulus ~ 1.2 ± 0.1 MPa) as probe. As soon as a defined compressive pre-load $F_L$ = 1.0 mN ± 0.1 mN is reached, the PDMS half-sphere is retracted. The pull-off force $F_{Ad}$ corresponds to the force minimum of the retraction part. The work of separation $W_{Se}$ corresponds to the shaded area enclosed by the retraction part and the zero force line.

**Force-displacement measurements**

$F_{Ad}$ and $W_{Se}$ were obtained by evaluation of the retraction parts of force-displacement curves (Figure 5). A sticky and compliant PDMS half-sphere is approached to the surface of the tested sample. As soon as contact forms, further displacement of the PDMS half-sphere towards the tested surface requires application of a compressive force with positive sign. As soon as a defined compressive pre-load $F_L$ is reached, the PDMS half-sphere is retracted. During retraction, adhesion causes the contact between PDMS half-sphere and tested surface to persist even beyond the point where the applied force becomes zero. The PDMS half-sphere remains attached to the tested surface until a negative pull-off force $F_{Ad}$ is reached, which corresponds to the force minimum in the retraction part of the corresponding force-displacement curve.

The work of separation $W_{Se}$ is the energy required to detach the PDMS half-sphere from the tested surface and corresponds to the area at negative forces enclosed by the zero force line and the retraction part of a force-displacement curve (Figure 5). Since the real contact area between the tested samples and the PDMS half-sphere is neither experimentally nor theoretically accessible, rather $W_{Se}$ than the work of adhesion ($W_{Se}$ divided by the contact area) is considered here. $W_{Se}$ nevertheless contains valuable information that can be qualitatively interpreted as long as the force-displacement measurements are carried out under the same conditions, such as the differences in $W_{Se}$ between FNMs and NMMAs, the differences in $W_{Se}$ in the presence and the absence of $TiO_2$ coatings as well as the differences in $W_{Se}$ in the presence and absence of mineral oil.

The previously reported force-displacement measurements on nanoporous fibrillar PS-*b*-P2VP adhesive pads optimized for strong adhesion were carried out with rigid sapphire spheres as probes.[36,37] However, this classical experimental layout appropriate for the measurement of high adhesion forces is not suitable for the detection of the small adhesion forces characteristic of the samples studied in this work. Therefore, we acquired force-displacement measurements using sticky and compliant viscoelastic PDMS half-spheres as probes, which are particularly suitable for the investigation of weakly adhesive surfaces.[19] The classical Johnson-Kendall-Roberts (JKR) model[45] for sphere-on-flat contacts assumes that the flat surface does not exhibit any kind of roughness. The surfaces of the NMMAs and FNMs studied here exhibit roughness on the 100 nm scale. Compliant PDMS half-spheres can adapt to the corrugated nanoporous surfaces, whereas rigid probes cannot. Hence, real contact area and the detected $F_{Ad}$ values will depend on the probe used. Consequently, the experimentally detected pull-off force $F_{Ad}$ is not equivalent to the pull-off force theoretically predicted by the JKR model, because the JKR model does not adequately describe the scenarios studied here. Also, the independence of $F_{Ad}$ on $F_L$ predicted by the JKR model does not always appropriately describe the experimental reality if the tested surfaces are rough. We found even using rigid spherical sapphire probes that $F_{Ad}$ on fibrillar adhesive pads depends on $F_L$ within a certain $F_L$ range.[46] Therefore, we normalized $F_{Ad}$ to $F_L$.

The force-displacement measurements were carried out in the presence or in the absence of mineral oil at the contact interface. The mineral oil was supplied from the undersides of the nanoporous PS-*b*-P2VP specimens – in the case of $TiO_2$-coated samples from the undersides opposite to the $TiO_2$-coated surfaces; in the case of NMMAs and $TiO_2$-NMMAs from the flat undersides opposite to the topographically patterned surfaces (Figure 1g and h). For this purpose, the undersides of the tested samples were placed on tissue impregnated with mineral oil. Because the nanopore networks of all tested samples were open, it is reasonable to assume the limiting case of a "drained" scenario, in which the deformation of the samples occurs at fixed pressure and in which the mineral oil can flow in or out of a deforming volume.

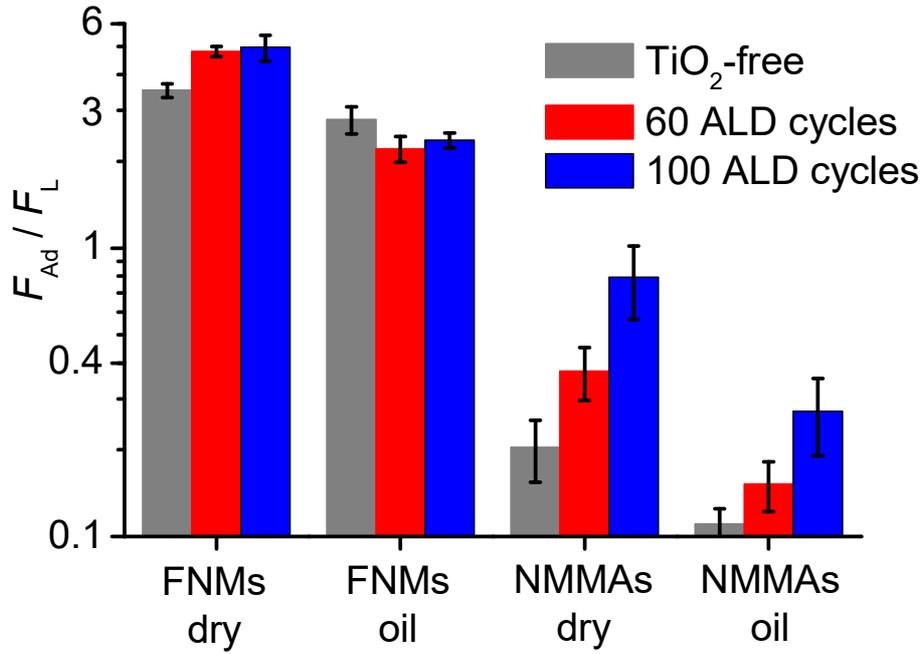

**Figure 6.** Pull-off force $F_{Ad}$ normalized to the loading force $F_L$ on FNMs and TiO$_2$-FNMs in the absence (flat dry) and the presence (flat oil) of mineral oil at the contact interface, as well as on NMMAs, TiO$_2$(60)-NMMAs and TiO$_2$(100)-NMMAs in the absence (NMMAs dry) and in the presence (NMMAs oil) of mineral oil at the contact interface. Note that the diagram is semi-logarithmic. Each bar represents the average of six measurements. The error bars denote standard deviations.

**Pull-off forces $F_{Ad}$**

We determined $F_{Ad}$ on FNMs and TiO$_2$-FNMs as well as on NMMAs, TiO$_2$(60)-NMMAs and TiO$_2$(100)-NMMAs (Figure 6) from the retraction parts of force-displacement curves (cf. Figure 5). We found the following trends, which we quantitatively evaluated by statistical analysis (degrees of freedom and $P$ values for pairwise comparisons dry/wet and uncoated/TiO$_2$(100) are listed in Tables S1 and S2 of the Supporting Information):

*1) Influence of topography.* NMMAs, TiO$_2$(60)-NMMAs and TiO$_2$(100)-NMMAs yielded $F_{Ad}/F_L$ values that were one order of magnitude smaller than those obtained on the corresponding FNMs and TiO$_2$-FNMs. This outcome was observed in the absence as well as in the presence of mineral oil at the contact interface.

*2) FNMs: influence of mineral oil.* In the presence of mineral oil, $F_{Ad}/F_L$ was in general smaller than in the absence of mineral oil. The set of all $F_{Ad}/F_L$ values obtained on uncoated and both types of TiO$_2$-FNMs in the absence of mineral oil exhibited highly significant differences from the corresponding set of $F_{Ad}/F_L$ values obtained in the presence of mineral oil (Kruskal-Wallis one way

ANOVA an Ranks, Dunn's Method; $H_{5,37}$ = 31.474, $P<0.001$). All pairwise comparisons of the $F_{Ad}/F_L$ values obtained on FNMs, TiO$_2$(60)-FNMs and TiO$_2$(100)-FNMs in the absence and in the presence of mineral oil revealed highly significant differences (t-test: P=0.005 and P<0.001).

*3) FNMs: influence of surface chemistry.* Without mineral oil, TiO$_2$-FNMs exhibited higher $F_{Ad}/F_L$ values than FNMs. The comparison of the set of $F_{Ad}/F_L$ values obtained on FNMs with the combined set of $F_{Ad}/F_L$ values obtained on TiO$_2$(60)-FNMs and TiO$_2$(100)-FNMs revealed that in the presence of TiO$_2$ $F_{Ad}/F_L$ was significantly higher (Kruskal-Wallis one way ANOVA on Ranks, Dunn's Method; $H_{2,16}$=8.159, $P$=0.017). The pairwise comparisons between the $F_{Ad}/F_L$ values obtained under dry conditions on FNMs as well as TiO$_2$(60)-FNMs and TiO$_2$(100)-FNMs using the *t*-test and the Mann-Whitney rank sum test ($P<0.05$) revealed that $F_{Ad}/F_L$ on FNMs was significantly lower than on TiO$_2$(60)-FNMs and TiO$_2$(100)-FNMs. However, no significant difference in $F_{Ad}/F_L$ was found between TiO$_2$(60)-FNMs and TiO$_2$(100)-FNMs ($P>0.05$).

The effect of the TiO$_2$ coatings was reversed in the presence of mineral oil; $F_{Ad}/F_L$ on FNMs tended to be higher than on TiO$_2$-FNMs. The comparison of the $F_{Ad}/F_L$ values obtained on FNMs and the combined sets of $F_{Ad}/F_L$ values obtained on TiO$_2$(60)- and TiO$_2$(100)-FNMs revealed significant differences (one way ANOVA: $F_{2,20}$=6.749, $P$=0.006). The pairwise comparison of the $F_{Ad}/F_L$ values obtained on the different samples in the presence of mineral oil revealed that $F_{Ad}/F_L$ on FNMs was significantly higher than on TiO$_2$(60)-FNMs (t-test: $P<0.05$). However, the comparisons of FNMs with either TiO$_2$(60)- or TiO$_2$(100)-FNMs did not reveal significant differences ($P>0.05$).

*4) NMMAs: influence of mineral oil.* Such as in the case of FNMs, $F_{Ad}/F_L$ on NMMAs was smaller in the presence than in the absence of mineral oil. The comparison of the combined set of the $F_{Ad}/F_L$ values obtained on NMMAs, TiO$_2$(60)-NMMAs and TiO$_2$(100)-NMMAs in the presence of mineral oil with that obtained in the absence of mineral oil revealed highly significant differences (Kruskal-Wallis one way ANOVA an Ranks: $H_{5,35}$=27.308, $P<0.001$). The pairwise comparison of TiO$_2$(60)-NMMAs and TiO$_2$(100)-NMMAs in the presence and the absence of mineral oil revealed a highly significant decrease in $F_{Ad}/F_L$ in the presence of mineral oil (t-test: $P<0.001$ and $P$=0.001). However, for uncoated NMMAs the differences were not significant (Mann-Whitney rank sum test: $P$=0.126).

*5) NMMAs: influence of surface chemistry.* In contrast to the results obtained on FNMs and TiO$_2$-FNMs, TiO$_2$ coatings increased $F_{Ad}/F_L$ on NMMAs both in the presence and in the absence of mineral oil at the contact interface. This effect was more pronounced for $TiO_2(100)$-NMMAs than for $TiO_2(60)$-NMMAs. Under dry conditions, the set of $F_{Ad}/F_L$ values obtained on uncoated NMMAs showed highly significant differences from the combined set of $F_{Ad}/F_L$ values obtained on $TiO_2(60)$- and $TiO_2(100)$-NMMAs (one way ANOVA: $F_{2,16}=12.920$, $P<0.001$). The pairwise comparison of the $F_{Ad}/F_L$ values obtained under dry conditions on NMMAs, $TiO_2(60)$-NMMAs and $TiO_2(100)$-NMMAs revealed that $F_{Ad}/F_L$ on uncoated NMMAs was significantly lower than on $TiO_2(60)$-NMMAs and $TiO_2(100)$-NMMAs (t-test and Mann-Whitney rank sum test: $P<0.05$). Moreover, $F_{Ad}/F_L$ on $TiO_2(60)$-NMMAs was significantly lower than on $TiO_2(100)$-NMMAs (t-test: $P<0.05$). In the presence of mineral oil the same trends were observed. $F_{Ad}/F_L$ increased from uncoated NMMAs to $TiO_2(60)$-NMMAs to $TiO_2(100)$-NMMAs. The set of $F_{Ad}/F_L$ values on uncoated NMMAs was significantly different from the combined set of $F_{Ad}/F_L$ values obtained on $TiO_2(60)$- and $TiO_2(100)$-NMMAs (Kruskal-Wallis one way ANOVA on ranks: $H_{2,19}=11.958$, $P=0.003$). All pairwise comparisons between the $F_{Ad}/F_L$ values obtained on NMMAs, $TiO_2(60)$-NMMAs and $TiO_2(100)$-NMMAs in the presence of mineral oil revealed significant differences (t-test and Mann-Whitney rank sum test: $P<0.05$).

*6) Minimization of $F_{Ad}$.* The lowest $F_{Ad}/F_L$ values of $0.11 \pm 0.01$ were obtained for NMMAs in the presence of mineral oil at the contact interface. For comparison, in the absence of mineral oil the $F_{Ad}/F_L$ values on FNMs and $TiO_2(100)$-FNMs amounted to $3.52 \pm 0.19$ and $4.97 \pm 0.51$.

**Work of separation $W_{Se}$**

The overall $W_{Se}$ range reached from $(1.0 \pm 0.3)*10^{-9}$ J for uncoated NMMAs in the absence of mineral oil to $(17.0 \pm 1.2)*10^{-8}$ J for $TiO_2(100)$-FNMs in the absence of mineral oil. Figure 7a displays $W_{Se}$ obtained on FNMs as well as on $TiO_2(60)$- and $TiO_2(100)$-FNMs, Figure 7b $W_{Se}$ obtained on NMMAs, $TiO_2(60)$-NMMAs and $TiO_2(100)$-NMMAs in the absence and in the presence of mineral oil. The following trends were apparent (degrees of freedom and $P$ values for pairwise comparisons dry/wet and uncoated/$TiO_2(100)$ are listed in Tables S3 and S4 of the Supporting Information):

*1) Influence of topography.* In the absence as well as in the presence of mineral oil, $W_{Se}$ obtained on NMMAs, $TiO_2(60)$-NMMAs and $TiO_2(100)$-NMMAs was one to two orders of magnitude smaller than on FNMs, $TiO_2(60)$- and $TiO_2(100)$-FNMs. Hence, the anti-adhesive behaviour of NMMAs, $TiO_2(60)$-NMMAs and $TiO_2(100)$-NMMAs is also apparent from the comparison of the corresponding $W_{Se}$ values. The differences in $W_{Se}$ between $TiO_2(60)$-FNMs and $TiO_2(60)$-NMMAs in the presence of mineral oil as well as between $TiO_2(100)$-FNMs and $TiO_2(100)$-NMMAs in the presence of mineral oil were highly significant (Kruskal-Wallis one way ANOVA on ranks: $H_{5,32}$=29.554, $P$<0.001). The differences in $W_{Se}$ between FNMs and NMMAs in the presence of mineral oil as well as between all sample pairs with different surface topography and the same surface chemistry under dry conditions were even more pronounced.

*2) FNMs: influence of mineral oil.* The pairwise comparisons of the sets of $W_{Se}$ values obtained on the same sample in the absence and the presence of mineral oil revealed that on FNMs $W_{Se}$ was significantly higher in the presence of mineral oil (t-test: $P$<0.05). However, on TiO$_2$(60)- and TiO$_2$(100)-FNMs $W_{Se}$ was significantly *smaller* in the presence of mineral oil (Mann-Whitney rank sum test: $P$<0.05 and t-test: $P$<0.001). Supply of mineral oil reduced $W_{Se}$ to 1/4 – 1/3 of the values obtained in the absence of mineral oil.

*3) FNMs: influence of surface chemistry in the absence of mineral oil.* Under dry conditions, $W_{Se}$ on FNMs was significantly smaller than on TiO$_2$(60)- and TiO$_2$(100)-FNMs. Coating of FNMs with TiO$_2$ was associated with an increase in $W_{Se}$ by a factor of 1.8 (60 ALD cycles) and of 2 (100 ALD cycles). The comparison of the set of $W_{Se}$ values obtained on FNMs with the combined sets of $W_{Se}$ values obtained on TiO$_2$(60)- and TiO$_2$(100)-FNMs revealed significant differences (Kruskal-Wallis one way ANOVA on ranks: $H_{2,15}$=9.983, $P$=0.007). In the absence of mineral oil, $W_{Se}$ on FNMs was significantly smaller than on TiO$_2$(60)-FNMs monoliths (t-test, $P$<0.05) and highly significantly smaller than on TiO$_2$(100)-FNMs (Mann-Whitney rank sum test, $P$<0.001). However, there was no significant difference between TiO$_2$(60)- and TiO$_2$(100)-FNMs (Mann-Whitney rank sum test: $P$>0.05).

*4) FNMs: influence of surface chemistry in the presence of mineral oil.* With mineral oil, $W_{Se}$ on FNMs was higher than on TiO$_2$(60)- and TiO$_2$(100)-FNMs. Moreover, $W_{Se}$ on TiO$_2$(60)-FNMs was significantly smaller than on TiO$_2$(100)-FNMs. All these differences were statistically significant (*t*-test and Mann-Whitney rank sum test: $P$<0.05). The comparison of the set of $W_{Se}$ values obtained on FNMs with the combined sets of $W_{Se}$ values obtained on TiO$_2$(60)- and TiO$_2$(100)-FNMs revealed highly significant differences (Kruscal-Wallis one way ANOVA on ranks: $H_{2,16}$=14.118, $P$<0.001).

*5) NMMAs: influence of mineral oil.* $W_{Se}$ tended to be higher in the presence than in the absence of mineral oil. Highly significant differences in $W_{Se}$ were found between the combined sets of $W_{Se}$ values obtained on NMMAs, TiO$_2$(60)-NMMAs and TiO$_2$(100)-NMMAs in the absence and in the presence of mineral oil (Kruskal-Wallis one way ANOVA an ranks: $H_{5,30}$=25.041, $P$<0.001).

However, the pairwise comparison of the sets of $W_{Se}$ values obtained on the same samples in the absence and the presence of mineral oil did not reveal significant differences (t-test: $P>0.05$).

*6) NMMAs: influence of surface chemistry in the absence of mineral oil.* Under dry conditions, $W_{Se}$ on TiO$_2$(60)-NMMAs was ~2.2 times higher and on TiO$_2$(100)-NMMAs ~7.8 times higher than on uncoated NMMAs. The set of $W_{Se}$ values on uncoated NMMAs was significantly different from the combined set of $W_{Se}$ values obtained on TiO$_2$(60)-NMMAs and TiO$_2$(100)-NMMAs (Kruskal-Wallis one way ANOVA on ranks: $H_{2,15}$=12.345, $P$=0.002). Without mineral oil, the t-test and the Mann-Whitney rank sum test indicated statistically significant differences ($P<0.05$) for all pairwise comparisons between the sets of $W_{Se}$ values obtained on uncoated NMMAs, TiO$_2$(60)-NMMAs and TiO$_2$(100)-NMMAs; the difference between the $W_{Se}$ values obtained on uncoated NMMAs and TiO2(60)-NMMAs was highly significant ($P<0.001$).

*7) NMMAs: influence of surface chemistry in the presence of mineral oil.* The trends observed in the presence of mineral oil were the same as those observed in the absence of mineral oil. In the presence of mineral oil, $W_{Se}$ on TiO$_2$(60)-NMMAs was ~3.2 times and $W_{Se}$ on TiO$_2$(100)-NMMAs ~9 times higher than on uncoated NMMAs. The difference between the set of $W_{Se}$ values measured on uncoated NMMAs and the combined set of $W_{Se}$ values obtained on TiO$_2$(60)-NMMAs and TiO$_2$(100)-NMMAs was highly significant (one way ANOVA: $F_{2,15}$=2.958, $P<0.001$). The pairwise comparison of the sets of $W_{Se}$ values obtained on NMMAs, TiO$_2$(60)-NMMAs and TiO$_2$(100)-NMMAs in the presence of mineral oil revealed significant differences for all possible combinations (t-test, $P<0.05$).

The differences between FNMs and NMMAs are obvious from a comparison of $W_{Se}$ in the absence of mineral oil at the contact interface (thereafter referred to as $W_{dry}$) and $W_{Se}$ in the presence of mineral oil at the contact interface (thereafter referred to as $W_{wet}$) (Figure 7c). $W_{dry}/W_{wet}$ amounted to 0.84 for NMMAs, to 0.96 for TiO$_2$(60)-NMMAs and to 0.73 for TiO$_2$(100)-NMMAs. $W_{dry}/W_{wet}$ amounted to 0.75 for FNMs, to 4.2 for TiO$_2$(60)-FNMs and to 3.3 for TiO$_2$(100)-FNMs.

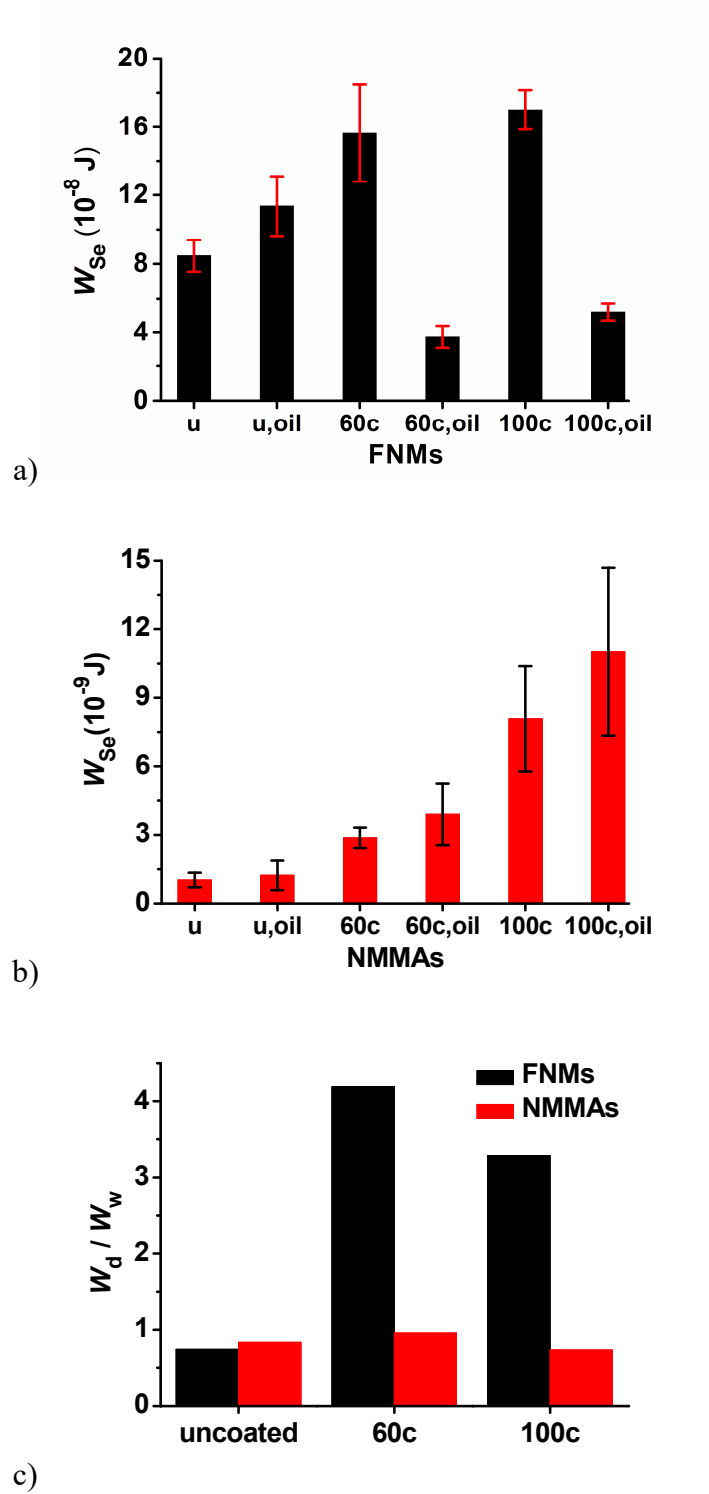

**Figure 7.** $W_{Se}$ determined by the evaluation of the retraction parts of force-displacement curves (cf. Figure 5). a) $W_{Se}$ on FNMs, TiO$_2$(60)-FNMs and TiO$_2$(100)-FNMs. b) $W_{Se}$ on NMMAs, TiO$_2$(60)-NMMAs and TiO$_2$(100)-NMMAs. Each bar in panels a) and b) represents the arithmetic mean value of six measurements. Standard deviations are indicated by error bars. c) Ratios $W_{dry}/W_{wet}$ of the average works of detachment in the absence ($W_{dry}$) and in the presence ($W_{wet}$) of mineral oil at the contact interface. "Uncoated" and "u" denote samples not coated with TiO$_2$, "60c" and "100c" denote samples coated with TiO$_2$ by 60 and 100 ALD cycles.

DISCUSSION

**Influence of topography**

Dry adhesion between two solid surfaces is reduced by the reduction of the real contact area. If the contacting counterpart surfaces are rigid, micro-roughness reduces the real contact area and consequently adhesion because contact is only formed at protrusions. The sticky and compliant surfaces of the PDMS half-spheres used here as probes for the force-displacement measurements can adapt to surface topographies characterized by small feature sizes below a few microns. Thus, on microsphere arrays with $r_s$ values in the submicron range contact area and adhesion are increased as compared to flat reference surfaces.[19] If the characteristic feature sizes of the surface topography are increased, the capability of compliant counterpart surfaces to adapt to the surface roughness decreases. Thus, the conformal contact between the two contacting surfaces breaks up and is replaced by islands of contact at protrusions surrounded by zones without contact. Real contact area and adhesion decrease and eventually fall below the values one would obtain on a flat reference surface. Then, surface roughness results in anti-adhesive behavior not only towards rigid but also towards soft counterpart surfaces.

On monolithic polystyrene (PS) microsphere arrays with $r_s$ values between 12.5 μm and 45 μm, $F_{Ad}/F_L$ measured with a compliant PDMS half sphere as probe amounted to only 10% to 16% of the value obtained on flat reference samples.[20] $F_{Ad}/F_L$ was most efficiently reduced on mixed monolithic PS microsphere arrays containing PS microspheres with $r_s$ values of 12.5 and 22.5 μm. Force-displacement measurements at several different positions reproducibly yielded $F_{Ad}/F_L$ values amounting to only 6 % of the value measured on flat reference samples; the standard deviation was even smaller than on PS microsphere arrays consisting of only one microsphere species.[20] This outcome was confirmed by force-displacement measurements on NMMAs at relative humidities of 2 % and 90 %.[21] Especially at relative humidities of 90 %, the mixed NMMAs studied here showed anti-adhesive properties that were even more pronounced than the anti-adhesive properties of their counterparts consisting of only one microsphere species with $r_s$ values of either 12.5 μm or 22.5 μm. The surface topography of the mixed NMMAs studied here can, therefore, be considered as hierarchical defect structure (similar to non-contiguous microsphere arrays), on which the actual contact area with the PDMS half-sphere is even more efficiently minimized than on surface topographies characterized by a single $r_s$ value. Under the conditions applied in this work, the topography-induced reduction in $F_{Ad}$ on NMMAs and TiO$_2$-NMMAs could be reproduced. In the absence of mineral oil, $F_{Ad}/F_L$ and $W_{Se}$ were one order of magnitude smaller than on the corresponding FNMs and TiO$_2$-FNMs (Figures 7 and 8). Hence, under dry conditions the anti-adhesive properties of mixed monolithic microsphere arrays with $r_s$ values of 12.5 and 22.5 μm were reproduced on non-polar PS,[20] on aprotic-polar P2VP, and on hydroxyl-terminated $TiO_2$ independent of the surface chemistry. It should be noted that ordered binary microsphere arrays[47] and ordered non-contiguous 2D microsphere arrays[48] reported in the literature are typically characterized by $r_s$ values ranging from a few 100 nm to a few microns. Therefore, for these surfaces rather adhesion enhancement than anti-adhesive behaviour towards compliant counterpart surfaces is to be expected.

**Influence of surface chemistry**

The increase in $F_{Ad}/F_L$ and $W_{Se}$ under dry conditions associated with ALD deposition of $TiO_2$, which is independent of the surface topography (flat, NMMAs), can be rationalized by assuming the formation of $TiO_2$-PDMS hydrogen bonds. The hydroxyl groups at the surface of the $TiO_2$ coatings can form hydrogen bonds with the backbone oxygen atoms that each PDMS repeat unit contains. In contrast, the surface of uncoated samples consists of aprotic P2VP blocks. Under dry conditions, adhesion between the PDMS half-spheres and the P2VP predominantly originates from van der Waals interactions since neither P2VP nor PDMS possess acidic hydrogen atoms, as required for the formation of hydrogen bonds. The importance of long-range van der Waals forces and contributions of subsurface layers to dry Gecko adhesion was stressed by Loskill et al.[49] The polarizability of a material, that is, the ease with which non-permanent dipole moments can be induced by external fields, crucially determines the strength of van der Waals interactions with this material. Adhesion on silicon wafers decreased along with increasing thickness of a silicon oxide layer covering the silicon, because the polarizability of silicon oxide is smaller than that of silicon. The systems used by us consist of a nanoporous PS-b-P2VP scaffold. Every repeat unit of the PS and P2VP blocks contains aromatic phenyl (PS) or pyridyl (P2VP) rings containing easily displaceable π atoms. Thus, it is reasonable to assume that the polarizability of PS-b-P2VP is higher than that of $TiO_2$. The contributions of the polarizable aromatic rings to the interfacial interactions should be reduced if they are buried below a dielectric $TiO_2$ layer. However, the increase in $F_{Ad}/F_L$ associated with the presence of $TiO_2$ coatings corroborates the notion that $TiO_2$-PDMS hydrogen bonds dominate interfacial interactions in the presence of $TiO_2$. The increase in $F_{Ad}/F_L$ and $W_{Se}$ along with increasing numbers of ALD cycles might be attributed to increased real contact areas related to thicker ALD coatings and to denser $TiO_2$ layers with higher area densities of hydroxyl groups.

**Influence of interfacial liquid on the pull-off force $F_{Ad}$**

For all NMMA species as well as for all types of FNMs studied here, $F_{Ad}/F_L$ tended to be smaller in the presence than in the absence of mineral oil (Figures 7 and 8). This outcome is in contrast to the results obtained on nanoporous fibrillar adhesive pads made of the same PS-b-P2VP block copolymer; on the nanoporous fibrillar adhesive pads, supply of mineral oil resulted in a pronounced *increase* in $F_{Ad}/F_L$ by one order of magnitude. This increase was ascribed to capillarity-supported formation of solid–solid contact.[36] Solid-solid contacts may form in the absence as well as in the presence of mineral oil. According to the modified Schargott-Popov-Gorb model, the probability of contact formation between nanorod-like contact elements and a counterpart surface influences $F_{Ad}/F_L$.[46] Capillarity-supported formation of solid-solid contact mediated by liquid bridges was assumed to increase this probability of contact formation and, therefore, $F_{Ad}/F_L$.[36] The bendability of the nanorod-like contact elements appears to be crucial to this effect. Capillarity-supported formation of solid–solid contact is, therefore, a peculiar feature of fibrillar nanoporous adhesive pads. In the case of FNMs and NMMAs, the absence of flexible topographic features with high aspect ratios prevents capillarity-supported formation of solid-solid contact, indicating that the impact of interfacial liquid on solid-solid contact formation differs for different surface topographies.

The reduction in $F_{Ad}/F_L$ on the samples studied in this work caused by the supply of mineral oil might be rationalized as follows. For a given contact geometry, changes in $F_{Ad}/F_L$ caused by the presence of liquids at contact interfaces depend on the Hamaker constants of the involved materials. In general, liquid supply may result in an increase or in a decrease in $F_{Ad}/F_L$. The observed decrease in $F_{Ad}/F_L$ corroborates the notion that the solid-solid interactions between sample surface and PDMS half-sphere are partially screened by the mineral oil. A second contribution to the reduction in $F_{Ad}/F_L$ could be related to poroelastic effects. The presence of mineral oil in the nanopores may reduce the deformability of the nanoporous PS-*b*-P2VP scaffold, which in turn would reduce the actual contact area with the PDMS half-sphere during force-displacement measurements. The mineral oil inside the nanopores would need to be displaced when the nanopores are squeezed by application of pressure. Since the nanopore systems of all tested samples are continuous and open, drainage of liquid away from the contact interface is, in principle, possible. In the case of the nanoporous fibrillar PS-*b*-P2VP adhesive pads studied in reference 37, the high surface-to-volume ratio of the nanoporous fibrillar contact elements enables highly efficient drainage of mineral oil in response of applied pressure; thus, the impact of poroelasticity may be reduced. However, the topographies of NMMAs and FNMs prevent such efficient drainage processes so that the impact of poroelastic stiffening might be more pronounced. On the other hand, there are also effects related to the presence of mineral oil that may even increase $F_{Ad}/F_L$. First, the elastic modulus of the PS-*b*-P2VP scaffold may be reduced by partial swelling of the PS-*b*-P2VP with mineral oil. However, the pronounced incompatibility of PS and P2VP[50] should prevent viscous deformation of the PS-*b*-P2VP scaffold by terminal flow. Secondly, capillarity-induced deformation of the involved surfaces[51-54] could counteract poroelastic stiffening. Yet, it is difficult to evaluate the contributions of these at least in part counteracting effects to the observed decrease in $F_{Ad}/F_L$ upon mineral oil supply. Evidently, the effects leading to increased $F_{Ad}/F_L$ values prevail for the scenarios considered here.

**Influence of interfacial liquid on the work of separation $W_{Se}$**

$W_{Se}$ is the second quantity that describes the separation of contacting counterpart surfaces. $W_{Se}$ includes the energy required to loosen the solid-solid contact between sample surface and PDMS half-sphere. In the presence of mineral oil, $W_{Se}$ also contains the energy required to stretch and rupture the liquid bridges between the contacting surfaces. This additional capillarity contribution arises from the Laplace pressure across the surfaces of the liquid bridges and from contact line tension.[55] However, the surfaces studied here are microscopically rough, corrugated and porous. The outermost P2VP blocks of the polymeric scaffold of uncoated samples might be partially swollen. Cohesion of the mineral oil inside the nanopore systems may play a role. Finally, it is unclear whether the liquid bridges are in mechanical equilibrium. Hence, it is challenging to quantitatively model the detachment of the PDMS half spheres. Nevertheless, the stretching and rupturing of the liquid bridges during the retraction of the PDMS half sphere influences only $W_{Se}$ as these phenomena take place after passing the force minimum at pull-off representing $F_{Ad}$.

On FNMs as well as on NMMAs and TiO$_2$-NMMAs supply of mineral oil is accompanied by an increase in $W_{Se}$. Hence, reductions of the solid-solid contributions related to possible screening of solid-solid interactions by mineral oil or reductions of $W_{Se}$ related to poroelastic effects in the presence of mineral oil are overbalanced by additional capillary contributions. However, on TiO$_2$-FNMs, $W_{Se}$ obtained without mineral oil is larger than $W_{Se}$ obtained in the presence of mineral oil. This difference to uncoated FNMs, NMMAs and TiO$_2$-NMMAs cannot be explained by poroelastic effects that would likewise be effective in the absence of TiO$_2$ coatings. Instead, on TiO$_2$-FNMs the decrease in the solid-solid contributions to $W_{Se}$ caused by mineral oil supply must be stronger than the increase in $W_{Se}$ related to capillarity. On the one hand, the solid-solid interactions between PDMS half-spheres and TiO$_2$-FNMs should be stronger than between PDMS half spheres and FNMs so that screening of solid-solid interactions by mineral oil has a more dramatic impact on TiO$_2$-FNMs. On the other hand, the absolute contact area between TiO$_2$-FNMs and the PDMS half-sphere within the contour of the contact circle is much larger than that between NMMAs and TiO$_2$-NMMAs. Thus, the importance of solid-solid interactions relative to capillarity is more pronounced on TiO$_2$-FNMs than on TiO$_2$-NMMAs.

**Interfacial liquid and surface chemistry**

In the presence of mineral oil, $F_{Ad}$ and $W_{Se}$ on TiO$_2$-FNMs tend to be smaller than on FNMs. This finding may be rationalized as follows. Taking into account that the surface of FNMs is corrugated, already small local deformations of the PS-*b*-P2VP scaffold may increase the contact area to the PDMS half-sphere. Such deformations may be facilitated by softening related to partial swelling of the PS-*b*-P2VP with mineral oil. Similar effects were previously reported for solid specimens; for example, softening of fresh Lotus leafs by moisture resulted in increased actual contact areas and higher adhesion as compared to rigid artificial analogues.[4] Moreover, moisture-induced softening resulted in adhesion enhancement on bioinspired fibrillar adhesive pads.[36] On the other hand, capillarity-induced surface deformation may result in better adhesion between soft and elastic surfaces than between rigid surfaces.[51-54] Despite the possible occurrence of softening-induced increase in the contact area, $F_{Ad}/F_L$ decreases if mineral oil is supplied. This behaviour may originate from other effects counteracting the softening-induced increase in the contact area, as discussed above. However, surface-induced softening may allow rationalizing the different behaviour of FNMs and TiO$_2$-FNMs apparent in the presence of mineral oil. The stiff TiO$_2$ coatings reduce the mechanical compliance of mineral oil-containing TiO$_2$-FNMs and suppress, therefore, softening of the surface of the TiO$_2$-FNMs. Hence, under wet conditions the actual contact areas between TiO$_2$-FNMs and PDMS half-spheres are smaller than the actual contact areas between FNMs and PDMS half-spheres. TiO$_2$-induced surface stiffening is evidently not balanced by the interactions between the TiO$_2$ surface hydroxyl groups and the PDMS half-spheres. This outcome supports the view that screening of interfacial solid-solid interactions between TiO$_2$-FNMs and PDMS half-spheres by mineral oil influences contact mechanics.

**Interfacial liquid and surface topography**

In contrast to the results obtained on FNMs, mineral oil-containing TiO$_2$-NMMAs show higher $F_{Ad}$ and $W_{Se}$ values than uncoated mineral oil-containing NMMAs. Obviously, the surface topography of the NMMAs and TiO$_2$-NMMAs weakens the effects that result in higher $F_{Ad}$ and $W_{Se}$ values on FNMs than on TiO$_2$-FNMs in the presence of mineral oil. Local softening of the NMMAs by mineral oil, as discussed above, does apparently not result in significant increases in the actual contact areas with the PDMS half-spheres as compared to dry NMMAs, because the NMMAs form only focal contacts with the PDMS half sphere at the caps of the microspheres. Stiffening of the NMMA surfaces by TiO$_2$ coatings does in turn not significantly reduce the actual contact areas with the PDMS half-spheres as compared to uncoated mineral-oil soaked NMMAs. On the other hand, screening of the TiO$_2$-PDMS solid-solid interactions by mineral oil is apparently less efficient on TiO$_2$-NMMAs than on TiO$_2$-FNMs. Possibly, this effect is related to the drainage of the mineral oil into the gaps between the TiO$_2$-NMMA microspheres. Thus, mineral oil might be depleted at the caps of the TiO$_2$-coated microspheres forming the contacts with the PDMS half-spheres. If one assumes that $F_{Ad}$ predominantly depends on the strength of the solid-solid contact, the higher $F_{Ad}$ values on mineral oil-soaked TiO$_2$-NMMAs as compared to mineral oil-soaked NMMAs could then be rationalized by strong TiO$_2$-PDMS interactions even in the presence of mineral oil. Finally, TiO$_2$-NMMAs have much smaller solid-solid contact areas with the PDMS half-spheres than TiO$_2$-FNMs. On mineral oil-soaked TiO$_2$-NMMAs the capillary contributions to $W_{Se}$ thus overcompensate possible reductions of the solid-solid contributions to $W_{Se}$ related to the presence of mineral oil.

CONCLUSIONS

We prepared nanoporous monoliths that allow delivery of liquids to their contact surfaces through continuous spongy nanopore systems with mean pore diameters of a few 10 nm. Thus, wet adhesion in the presence of liquid at the contact surface could deliberately be exploited to tailor contact mechanics. We have comparatively investigated samples with aprotic polymeric or titania surfaces that had either macroscopically flat surface topography (FNMs) or surface topographies characterized by arrays of microspheres with diameters of a few 10 μm (NMMAs). Adhesion under dry and wet conditions was investigated by force-displacement measurements using sticky probes, an experimental technique optimized for the characterization of weakly adhesive surfaces. On all samples, the pull-off forces $F_{Ad}$ were smaller under wet than under dry conditions. Possible origins of the reduction of $F_{Ad}$ by mineral oil supply may include screening of solid-solid interactions by the liquid or poroelastic effects. The impact of interfacial liquids on the work of separation $W_{Se}$ is more complex, since $W_{Se}$ is also influenced by capillarity. Under dry conditions, we obtained smaller $F_{Ad}$ and $W_{Se}$ values on FNMs with polymeric surface than on TiO$_2$-FNMs, because the latter form hydrogen bonds to the sticky probes. Under wet conditions, this trend was reversed; $F_{Ad}$ and $W_{Se}$ on TiO$_2$-FNMs were smaller than on FNMs with aprotic polymeric surface. This result may be rationalized by the suppression of deformations related to liquid-induced softening, which increase the actual contact area and thus adhesion, by stiff TiO$_2$ coatings. In this specific case, the enhanced interactions between TiO$_2$ and the sticky probes did not balance the stiffening-induced decrease in adhesion, possibly because of partial screening of interfacial solid-solid interactions by interfacial liquid. Moreover, $W_{Se}$ on TiO$_2$-FNMs drastically *decreased* under wet conditions as compared to dry conditions. Hence, under wet conditions the additional capillary contributions to $W_{Se}$ did not compensate the screening of interfacial solid-solid interactions by the liquid. On NMMAs and TiO$_2$-NMMAs, $F_{Ad}$ and $W_{Se}$ were reduced by one order of magnitude as compared to their flat nanoporous counterparts. Thus, NMMAs and TiO$_2$-NMMAs showed anti-adhesive properties.

Furthermore, for NMMAs and TiO$_2$-NMMAs the impact of liquid-induced surface deformation and screening of interfacial solid-solid interactions by liquid on adhesion was drastically reduced. In contrast to FNMs and TiO$_2$-FNMs, we obtained under dry as well as under wet conditions on TiO$_2$-NMMAs always larger $F_{Ad}$ and $W_{Se}$ values than on NMMAs. Moreover, $W_{Se}$ on NMMAs and TiO$_2$-NMMAs increased upon supply of liquid because additional capillary contributions overcompensated any liquid-induced decrease in interfacial solid-solid interactions.

$F_{Ad}$ and $W_{Se}$ can be varied by at least one order of magnitude by simple topographic and chemical modifications of the surfaces of the nanoporous PS-*b*-P2VP monoliths combined with delivery of liquid to the contact interfaces through the nanopores. The results reported here may contribute to the development of rational design principles for functional surfaces exploiting deployment or drainage of interfacial liquids as well as liquid-induced softening for adhesion management. Examples for liquid-mediated adhesion management may include liquid-triggered contact loosening or adhesion enhancement by drainage of undesired interfacial liquid away from contact surfaces taking advantage of the presence of continuous nanopore systems. The results reported here may also help improve our understanding of the contact mechanics of biological surfaces. It is known that under wet conditions insect adhesion is strongly reduced if compared to dry conditions.[32] This can be explained by a reduction of the interfacial solid-solid interactions and by a reduction of capillary interactions due to the absence of air-liquid-solid interfaces. It is also known that insect adhesion is reduced along with increasing amount of fluid.[56,57] As shown above, similar effects were obtained on the non-biological nanoporous samples investigated here.

MATERIALS AND METHODS

**Preparation of flat nanoporous PS-*b*-P2VP monoliths.** Asymmetric PS-*b*-P2VP ($M_n$ (PS) = 101 000 g/mol; $M_n$(P2VP) = 29 000 g/mol; $M_w/M_n$ = 1.60; volume fraction of P2VP 21 %; bulk period ~51 nm) was obtained from Polymer Source Inc., Canada. 0.2 g PS-*b*-P2VP was dissolved in 2 ml THF (99.9 %, Sigma Aldrich) at 40°C for 1 h. The solution was then kept at room temperature for 24 h to remove bubbles and poured onto flat PDMS substrates. The THF was slowly evaporated at room temperature for 1 week by placing the samples in Petri dishes covered with glass lids. Subsequently, the samples were dried for 24 h under vacuum at room temperature. Finally, the PDMS substrates were detached and flat solid PS-*b*-P2VP specimens were obtained. About 500 μm thick FNMs were formed by swelling-induced pore generation in ethanol at 60°C for 4 h.

**Preparation of NMMAs.** PDMS secondary molds were prepared as described elsewhere.[20] Solutions containing 200 mg PS-*b*-P2VP per 1.2 mL THF (99,9 %, Sigma Aldrich) were heated to 40° C for 1 h, kept at room temperature for 24 h to remove air bubbles and then deposited onto the PDMS molds. To slowly evaporate the THF, the samples were placed in Petri dishes covered with glass lids for 1 week at room temperature and then dried for 24 h under vacuum at room temperature. Detachment from the PDMS molds yielded monolithic specimens decorated with PS-*b*-P2VP microsphere arrays tightly connected to ~500 μm thick PS-*b*-P2VP substrates over areas of 5 mm x 5 mm. NMMAs were then obtained by swelling-induced pore generation in ethanol at for 4 h at 60°C. The PDMS molds were reused multiple times.

**ALD of $TiO_2$.** Adapting a previously reported method,[38] ALD was performed in a Delta f-100-31 reactor (Wuxi MINT micro- and nanotech Co., China) at a temperature of 80°C. $TiCl_4$ and deionized $H_2O$ were used as precursors. The flow rate of $N_2$ was adjusted to 20 sccm. $TiO_2$ was deposited by repetitions of the following cycle: $TiCl_4$/ET/$N_2$/$H_2O$/ET/$N_2$ = 15ms/50s/10s/30ms/50s/10s, where ET is the exposure time. Typically, 1.1 Å $TiO_2$ were deposited per cycle. The thickness of $TiO_2$ films deposited as reference samples onto cleaned and dried (100) Si wafers covered by a native $SiO_2$ layer (Table S5) was measured using a spectroscopic ellipsometer EASE M-2000U (J.A. Woollam Co. Inc., USA).

**Scanning electron microscopy (SEM).** SEM investigations were carried out on a Hitachi S4800 microscope (Hitachi High-Technologies Corporation, Japan) as well as on a Zeiss Auriga SEM operated at an accelerating voltage of 5 kV. The investigated sample surfaces were coated with Au-Pd alloy before imaging.

**Force-displacement measurements**

For the experiments reported in this work, a set of PDMS half-spheres with a diameter of 3 mm was simultaneously prepared under exactly the same conditions. Sylgard Elastomer 184 (Dow Corning) was used as PDMS prepolymer formulation. Base and curing agent were thoroughly mixed at a weight ratio of 8:1 for 3 min using steel spatula. The obtained PDMS prepolymer formulation was kept under ambient conditions until all air bubbles had vanished and then poured onto polyvinylsiloxan (PVS) molds containing several half-spherical cavities that had in turn been obtained by replication molding of sapphire spheres glued on a glass slide.[19] Subsequently, the PDMS prepolymer formulation was cured for 2 h at 60°C. Composition of the prepolymer formulation, curing conditions and post-curing storage conditions were identical for all PDMS half-spheres used in this work. For each sample, a new PDMS half-sphere from the set of identical PDMS half-spheres was used. Force-displacement measurements on NMMAs and TiO$_2$-NMMAs with areas of about 0.5 x 0.5 cm$^2$ as well as on FNMs and TiO$_2$-FNMs were carried out at a temperature of 28 °C and a relative humidity of 42.6 % with a home-made microforce biotester Basalt-01 (TETRA GmbH, Ilmenau, Germany). The Microforce tester Basalt-01 used a distance-controlled feedback loop. The force was controlled by the displacement. The PDMS half-spheres were mounted on a spring with a spring constant of 203.9 Nm$^{-1}$ and displaced towards the sample surfaces until a positive loading force $F_L$ = 1±0.1 mN was reached. Immediately after the preset $F_L$ value had been reached, the PDMS half-spheres were retracted. The approach and retraction speeds of the PDMS half-spheres were 100 µm/s. Series of force-displacement measurements on a specific sample were taken as follows. At first, the flat underside of the sample was placed on tissue paper with an area of about 1 cm$^2$, and a series of force-displacement measurements was carried out under dry conditions. Then, 5 µL mineral oil were dropped onto the uncovered part of the tissue paper, and we acquired a series of force-displacement curves under wet but otherwise identical conditions without moving the sample and without changing the PDMS half-sphere. We used mineral oil purchased from Sigma-Aldrich with a viscosity of 1.42-1.70 x 10$^{-5}$ m$^2$s$^{-1}$ at 40°C, as specified by the supplier. The wait time between mineral oil supply and the measurement of the first force-displacement curve after mineral oil supply was 30 min. This time period is sufficient to make sure that the tested samples were saturated with mineral oil.[37] Prior to any measurement after mineral oil supply, the PDMS half-spheres were cleaned with tissue paper. $W_{Se}$ was numerically estimated from the retraction parts of the force-displacement curves using MATLAB software (MathWorks, Inc.) by applying the trapezoidal rule (Newton-Cotes formula).[58] Although not applicable to viscoelastic materials,[59] we used the JKR theory to determine the effective elastic module of the system "PDMS half-sphere"/"FNM", but not to determine the work of adhesion. The reason for this is that the viscosity of the considered materials only weakly influences the elastic modulus but strongly influences the work of adhesion. The effective elastic modulus of the system "PDMS half-sphere"/"FNM" amounted to 1.2 ± 0.1 MPa and was calculated by evaluation of 10 force-displacement measurements on flat PS-*b*-P2VP monoliths using the JKR theory.[45] Equation 9 in reference 60 was used to fit the force-displacement curves. The exact displacement at which the PDMS half-sphere formed contact with the tested FNMs was a fit parameter that was not further used.

AUTHOR INFORMATION

Address correspondence to lxue@uos.de (L. Xue), yongwang@njtech.edu.cn (Y. Wang), sgorb@zoologie.uni-kiel.de (S. N. Gorb), martin.steinbart@uos.de (M. Steinhart).

**Notes**

The authors declare no competing financial interest.

ACKNOWLEDGEMENTS

Support by the German Research Foundation (DFG Priority Program 1420 "Biomimetic Materials Research: Functionality by Hierarchical Structuring of Materials") and by the European Research Council (ERC-CoG-2014; project 646742 INCANA) is gratefully acknowledged. A. Eichler-Volf, H. Chen, Y. Wang and M. Steinhart thank the Alexander von Humboldt Foundation for support through the Research Group Linkage Program.

SUPPORTING INFORMATION

**Figure S1:** SEM images of NMMAs. **Figure S2:** a) Frequency density of pore sizes and cumulative volume distribution as function of NMMA pore size; b) representative nitrogen sorption isotherm of a NMMA. **Table S1:** degrees of freedom and $P$ values of the pairwise comparisons $F_{Ad}/F_L$ dry vs. wet. **Table S2:** degrees of freedom and $P$ values of the pairwise comparisons $F_{Ad}/F_L$ uncoated vs. 100c. **Table S3:** degrees of freedom and $P$ values of the pairwise comparisons $W_{Se}$ dry vs. wet. **Table S4:** degrees of freedom and $P$ values of the pairwise comparisons $W_{Se}$ uncoated vs. 100c. **Table S5**: Dependence of the thickness of the deposited $TiO_2$ layers on the number of ALD cycles.

REFERENCES

1. Boesel, L. F.; Greiner, C.; Arzt, E.; del Campo, A. Gecko-Inspired Surfaces: A Path to Strong and Reversible Dry Adhesives *Adv. Mater.* **2010**, *22***,** 2125-2137.
2. Kwak, M. K.; Pang, C.; Jeong, H. E.; Kim, H. N.; Yoon, H.; Jung, H. S.; Suh, K. Y. Towards the Next Level of Bioinspired Dry Adhesives: New Designs and Applications. *Adv. Funct. Mater.* **2011**, *21*, 3606-3616.
3. Persson, B. N. J. On the Theory of Rubber Friction. *Surf. Sci.* **1998**, *401*, 445-454.
4. Koch K.; Bhushan, B.; Jung, Y. C.; Barthlott W. Fabrication of Artificial Lotus Leaves and Significance of Hierarchical Structure for Superhydrophobicity and Low Adhesion. *Soft Matter* **2009**, *5*, 1386–1393.
5. Shirtcliffe, N. J.; McHale, G.; Newton, M. I. The Superhydrophobicity of Polymer Surfaces:

Recent Developments. *J. Polym. Sci. Part B Polym. Phys.* **2011**, *49*, 1203-1217.

6. Bhushan, B.; Jung, Y. C. Natural and Biomimetic Artificial Surfaces for Superhydrophobicity, Self-Cleaning, Low Adhesion, and Drag Reduction. *Prog. Mater. Sci.* **2011**, *56*, 1–108.

7. Yan, Y. Y.; Gao, N.; Barthlott, W. Mimicking Natural Superhydrophobic Surfaces and Grasping the Wetting Process: A Review on Recent Progress in Preparing Superhydrophobic Surfaces. *Adv. Colloid Interface Sci.* **2011**, *169*, 80–105.

8. Liu, K.; Jiang L. Bio-Inspired Design of Multiscale Structures for Function Integration. *Nano Today* **2011**, *6*, 155-175.

9. Nishimoto, S.; Bhushan, B. Bioinspired Self-cleaning Surfaces with Superhydrophobicity, Superoleophobicity, and Superhydrophilicity. *RSC Adv.* **2013**, *3*, 671–690.

10. Butt, H.-J.; Roisman, I. V.; Brinkmann, M.; Papadopoulos, P.; Vollmer, D.; Semprebon, C. Liquid-Repellent Surfaces Minimizing Solid/Liquid Adhesion: Characterization of Super Liquid-Repellent Surfaces. *Curr. Opin. Colloid Interface Sci.* **2014**, *19*, 343-355.

11. Badge, I.; Bhawalkar, S. P.; Jia, L.; Dhinojwala, A. Tuning Surface Wettability Using Single Layered and Hierarchically Ordered Arrays of Spherical Colloidal Particles. *Soft Matter* **2013**, *9*, 3032–3040.

12. Mertaniemi, H.; Laukkanen, A.; Teirfolk, J.-E.; Ikkala, O.; Ras, R. H. A. Functionalized Porous Microparticles of Nanofibrillated Cellulose for Biomimetic Hierarchically Structured Superhydrophobic Surfaces. *RSC Adv.* **2012**, *2*, 2882–2886.

13. Ebert, D.; Bhushan, B. Durable Lotus-Effect Surfaces with Hierarchical Structure Using Micro- and Nanosized Hydrophobic Silica Particles. *J. Colloid Interface Sci.* **2012**, *368*, 584–591.

14. Raza, M. A.; Kooij, E. S.; van Silfhout, A.; Zandvliet, H. J.W.; Poelsema B. A Colloidal Route to Fabricate Hierarchical Sticky and Non-Sticky Substrates. *J. Colloid Interface Sci*. **2012**, *385*, 73–80.

15. Wang, L.; Wen, M.; Zhang, M.; Jiang, L.; Zheng, Y. Ice-Phobic Gummed Tape with Nano-Cones on Microspheres. *J. Mater. Chem. A* **2014**, *2*, 3312–3316.

16. Ho, A. Y. Y.; Van, E. L.; Lim, C. T.; Natarajan, S.; Elmouelhi, N.; Low, H. Y.; Vyakarnam, M.; Cooper, K.; Rodriguez, I. Lotus Bioinspired Superhydrophobic, Self-Cleaning Surfaces from Hierarchically Assembled Templates. *J. Polym. Sci. Part B Polym. Phys.* **2014**, *52*, 603-609.

17. Karaman, M.; Cabuk, N.; Özyurt, D.; Köysüren, Ö. Self-Supporting Superhydrophobic Thin Polymer Sheets that Mimic the Nature's Petal Effect. *Appl. Surface Sci.* **2012**, *259*, 542– 546.

18. Zhang, L.; Zhang, X.; Dai, Z.; Wua, J.; Zhao, N.; Xu, J. Micro–Nano Hierarchically Structured Nylon 6,6 Surfaces with Unique Wettability. *J. Colloid Interface Sci*. **2010**, *345*, 116–119.

19. Purtov, J.; Gorb, E. V.; Steinhart, M.; Gorb, S. N. Measuring of the Hardly Measurable: Adhesion Properties of Anti-adhesive Surfaces. *Appl. Phys. A* **2013**, *111*, 183–189.

20. Eichler-Volf, A.; Kovalev, A.; Wedeking, T.; Gorb, E. V.; Xue, L.; You, C.; Piehler, J.; Gorb, S. N.; Steinhart, M. Bioinspired Monolithic Polymer Microsphere Arrays as Generically Anti-Adhesive Surfaces. *Bioinspiration Biomimetics* **2016**, *11*, 025002.

21. Eichler-Volf, A.; Xue, L.; Kovalev, A.; Gorb, E. V.; Gorb, S. N.; Steinhart, M. Nanoporous Monolithic Microsphere Arrays Have Anti-Adhesive Properties Independent of Humidity. *Materials* **2016**, *9*, 373.

22. Mishchenko, L.; Hatton, B.; Bahadur, V.; Taylor, J. A.; Krupenkin, T.; Aizenberg, J. Design of Icefree Nanostructured Surfaces Based on Repulsion of Impacting Water Droplets. *ACS Nano* **2010**, *4*, 7699–7707.

23. Wong, T. S.; Kang, S. H.; Tang, S. K. Y.; Smythe, E. J; Hatton, B. D.; Grinthal, A.; Aizenberg, J. Bioinspired Self-Repairing Slippery Surfaces with Pressure-Stable Omniphobicity. *Nature* **2011**, *477*, 443-447.

24. Kim, P.; Wong, T.-S.; Alvarenga, J.; Kreder, M. J.; Adorno-Martinez, W. E.; Aizenberg, J. Liquid-Infused Nanostructured Surfaces with Extreme Anti-Ice and Anti-Frost Performance. *ACS Nano* **2012**, *6*, 6569-6577.

25. Bauchhenss, E. Die Pulvillen von *Calliphora erythrocephala* (Diptera, Brachycera) als Adhäsionsorgane. *Zoomorphologie* **1979**, *93*, 99-123.

26. Dirks, J.-H.; Federle, W. Mechanisms of Fluid Production in Smooth Adhesive Pads of Insects. *J. R. Soc., Interface* **2011**, *8*, 952–960.

27. Gorb, S. N. The Design of the Fly Adhesive Pad: Distal Tenent Setae are Adapted to the Delivery of an Adhesive Secretion. *Proc. R. Soc. B* **1998**, *265*, 747–752.

28. Gorb, S. N. Porous Channels in the Cuticle of the Head-Arrester System in Dragon/Damselflies. *Microsc. Res. Techn.* **1997**, *37*, 583–591.

29. Gorb, S. N.; Niederreger, S.; Hayashi, C. Y.; Summers, A. P.; Vötsch, W.; Walther, P. Silk-Like Secretion from Tarantula Feet. *Nature* **2006,** 443, 407.

30. Eisner, T.; Aneshansley, D. J. Defense by Foot Adhesion in a Beetle (*Hemisphaerota Cyanea*). *Proc. Natl. Acad. Sci. USA* **2000**, *97*, 6568-6573.

31. Federle, W.; Riehle, M.; Curtis, A. S. G.; Full R. J. An Integrative Study of Insect Adhesion: Mechanics and Wet Adhesion of Pretarsal Pads in Ants. *Integr. Comp. Biol.* **2002**, *42*, 1100–1106.

32. Langer, M. G.; Ruppersberg, J. P.; Gorb, S. N. Adhesion Forces Measured at the Level of a Terminal Plate of the Fly's Seta. *Proc. R. Soc. B* **2004**, *271*, 2209–2215.

33. Gorb, E. V.; Hosoda, N.; Miksch, C.; Gorb, S. N. Slippery Pores: Anti-Adhesive Effect of

Nanoporous Substrates on the Beetle Attachment System. *J. R. Soc. Interface* **2010**, *7*, 1571–1579.

34. Dirks, J.-H.; Federle, W. Fluid-Based Adhesion in Insects – Principles and Challenges. *Soft Matter* **2011**, *7*, 11047–11053.

35. Drotlef, D.-M.; Stepien, L.; Kappl, M.; Barnes, W. J. P.; Butt, H.-J.; Del Campo, A. Insights Into the Adhesive Mechanisms of Tree Frogs using Artificial Mimics *Adv. Funct. Mater.* **2013**, *23*, 1137–1146.

36. Xue, L.; Kovalev, A.; Dening, K.; Eichler-Volf, A.; Eickmeier, H.; Haase, M.; Enke, D.; Steinhart, M.; Gorb, S. N. Reversible Adhesion Switching of Porous Fibrillar Adhesive Pads by Humidity. *Nano Lett.* **2013**, *13*, 5541-5548.

37. Xue, L.; Kovalev, A.; Eichler-Volf, A.; Steinhart, M.; Gorb S. N. Humidity-Enhanced Wet Adhesion on Insect-Inspired Fibrillar Adhesive Pads. *Nat. Commun.* **2015**, *6*, 6621.

38. Li, F.; Yao, X.; Wang, Z.; Xing, W.; Jin, W.; Huang, J.; Wang, Y. Highly Porous Metal Oxide Networks of Interconnected Nanotubes by Atomic Layer Deposition. *Nano Lett.* **2012**, *12*, 5033-5038.

39. Wang, Y.; Gösele, U.; Steinhart, M. Mesoporous Block Copolymer Nanorods by Swelling-Induced Morphology Reconstruction. *Nano Lett.* **2008**, *8*, 3548-3553.

40. Wang, Y.; He, C.; Xing, W.; Li, F.; Tong, L.; Chen, Z.; Liao, X.; Steinhart, M. Nanoporous Metal Membranes with Bicontinuous Morphology from Recyclable Block-Copolymer Templates. *Adv. Mater.* **2010**, *22*, 2068-2072.

41. Wang, Y.; Li, F. An Emerging Pore-Making Strategy: Confined Swelling-Induced Pore Generation in Block Copolymer Materials. *Adv. Mater.* **2011**, *23*, 2134-2148.

42. Yao, X.; Guo, L.; Chen, X.; Huang, J.; Steinhart, M.; Wang, Y. Filtration-Based Synthesis of Micelle-Derived Composite Membranes for High-Flux Ultrafiltration. *ACS Appl. Mater. Interfaces* **2015**, *7*, 6974−6981.

43. Nam, H. J.; Jung, D.-Y.; Yi, G.-R.; Choi, H. Close-Packed Hemispherical Microlens Array from Two-Dimensional Ordered Polymeric Microspheres. *Langmuir* **2006**, *22,* 7358-7363.

44. Rengarajan, G. T.; Walder, L.; Gorb, S. N.; Steinhart M. High-Throughput Generation of Micropatterns of Dye-Containing Capsules Embedded in Transparent Elastomeric Monoliths by Inkjet Printing. *ACS Appl. Mater. Interfaces* **2012**, *4*, 1169-1173.

45. Johnson, K. L.; Kendall, K.; Roberts, A. D. Surface Energy and the Contact of Elastic Solids. *Proc. R. Soc. A* **1971**, *324,* 301-313.

46. Xue, L.; Kovalev, A.; Thöle, F.; Rengarajan, G. T.; Steinhart, M.; Gorb, S. N. Tailoring Normal Adhesion of Arrays of Thermoplastic, Spring-Like Polymer Nanorods by Shaping Nanorod Tips. *Langmuir* **2012**, *28*, 10781–10788.

47. Kitaev, V.; Ozin, G. A. Self-Assembled Surface Patterns of Binary Colloidal Crystals. *Adv. Mater.* **2003**, *15*, 75–78.

48. Choi, H. K.; Kim, M. H.; Im, S. H.; Park, O. O. Fabrication of Ordered Nanostructured Arrays Using Poly(dimethylsiloxane) Replica Molds Based on Three-Dimensional Colloidal Crystals. *Adv. Funct. Mater.* **2009**, *19*, 1594–1600.

49. Loskill, P.; Puthoff, J.; Wilkinson, M.; Mecke, K.; Jacobs, K.; Autumn, K. Macroscale Adhesion of Gecko Setae Reflects Nanoscale Differences in Subsurface Composition. *J. R. Soc. Interface* *2012*, **10**, 20120587.

50. Schulz, M. F.; Khandpur, A. K.; Bates, F. S.; Almdal, K.; Mortensen, K.; Hajduk, D. A.; Gruner, S. M. Phase Behavior of Polystyrene-Poly(2-vinylpyridine) Diblock Copolymers. *Macromolecules* **1996**, *29*, 2857-2867.

51. Xue, X.; Polycarpou, A. A. Meniscus Model for Noncontacting and Contacting Sphere-on-Flat Surfaces Including Elastic-Plastic Deformation. *J. Appl. Phys.* **2008**, *103*, 023502.

52. Butt, H.-J.; Barnes, W. J. P.; Del Campo, A.; Kappl, M.; Schönfeld, F. Capillary Forces Between Soft, Elastic Spheres. *Soft Matter* **2010**, *6*, 5930–5936.

53. Wexler, J. S.; Heard, T. M.; Stone, H. A. Capillary Bridges Between Soft Substrates. *Phys. Rev. Lett.* **2014**, *112*, 066102.

54. Li, K.; Cai, S. Wet Adhesion Between Two Soft Layers. *Soft Matter* **2014**, *10*, 8202–8209.

55. Vogel, M. J.; Steen, P. H. Capillarity-Based Switchable Adhesion. *Proc. Natl. Acad. Sci. USA* **2010**, 107, 3377-3381.

56. Drechsler, P.; Federle, W. Biomechanics of Smooth Adhesive Pads in Insects: Influence of Tarsal Secretion on Attachment Performance. *J. Comp. Physiol. A* **2006**, *192*, 1213-1222.

57. Labonte, D.; Federle, W. Rate-Dependence of 'Wet' Biological Adhesives and the Function of the Pad Secretion in Insects. *Soft Matter* **2015**, *11*, 8661-8673.

58. Atkinson, K. E. Numerical Integration. In *An Introduction to Numerical Analysis*; Atkinson, K. E., Ed.; Wiley: Hoboken, NJ, 1988; 2nd ed., pp 251-263.

59. Lin, Y.-Y.; Hui, C. Y., Mechanics of Contact and Adhesion between Viscoelastic Spheres: An Analysis of Hysteresis during Loading and Unloading. *J. Polym. Sci. Part B Polym. Phys.* **2002**, *40*, 772–793.

60. Ebenstein, D. M.; Wahl, K. J. A Comparison of JKR-Based Methods to Analyze Quasi-Static and Dynamic Indentation Force Curves. *J. Colloid Interface Sci.* **2006**, *298*, 652-662.

TOC graphic

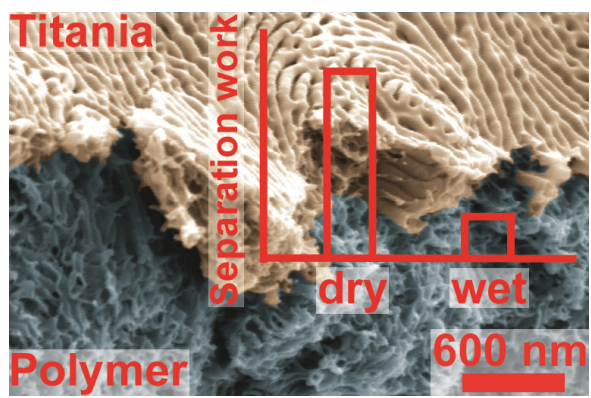